\documentclass{ws-procs9x6}

\begin{document}

\title{THE PION FORM FACTOR IN AdS/QCD}

\author{HERRY J. KWEE$^*$ and RICHARD F. LEBED$^\dagger$}

\address{Department of Physics, Arizona State University,\\
Tempe, AZ 85287-1504, USA\\
$^*$E-mail: Herry.Kwee@asu.edu\\
$^\dagger$E-mail: Richard.Lebed@asu.edu\\
http://phy.asu.edu}

\begin{abstract}
Holographic QCD provides a unique framework in which to compute QCD
observables.  In this talk we summarize recent numerical work on
computing the pion electromagnetic form factor using an AdS/QCD action
that includes both spontaneous and explicit chiral symmetry breaking.
We consider both hard- and soft-wall model results and develop an
intermediate background that supports the best features of both.  We
also begin to see possible evidence in the fit for the presence of
$1/N_c$ corrections.
\end{abstract}


\bodymatter

\section{Introduction and Background}\label{intro}

Studies of the duality between strongly-coupled Yang-Mills gauge
theories and weakly-coupled gravity on curved backgrounds, originating
with the anti-de~Sitter/conformal field theory (AdS/CFT)
correspondence\cite{AdSCFT}, have become prominent in studies of
strongly-coupled field theories.  In QCD, the most accessible of these
theories, the approach is dubbed ``AdS/QCD.''  In contrast to the
exact conformality of the original AdS/CFT example ${\cal N} \! = \!
4$ SUSY, the approximate conformality of QCD is broken by explicit
mass scales such as $\Lambda_{\rm QCD}$ and quark masses (as evidenced
by confinement and chiral symmetry breaking), which must in some
manner be incorporated into the theory if one hopes to achieve a
satisfactory picture for the rich spectrum and dynamics of QCD.

Since a great deal of AdS/QCD phenomenology has already been studied,
even in just the meson
sector\cite{KKSS,EKSS,BoschiFilho:2002vd,deTeramond:2005su,
Evans:2006ea,Colangelo:2007pt,Forkel:2007cm,DaRold:2005zs,
Hirn:2005nr,Ghoroku:2005vt,HuangZuo,heavyquark,Hambye:2005up,GR1,GR2,
Hong:2004sa,Radyushkin:2006iz,BdT,GR3,GR4,Colangelo:2008us,
dePaula:2008fp,Agaev:2008uj,Zeng:2008sx,Batell:2008zm,KL1,KL2}, we
provide only a summary of the most salient features of the approach,
leading eventually to a discussion of our work on the pion
electromagnetic form factor $F_\pi (Q^2)$\cite{KL1,KL2}.  In the
holographic approach one begins with the 5-dimensional AdS ``sliced''
metric,
\begin{equation}
ds^2 = g^{\vphantom\dagger}_{MN} \, dx^M dx^N = \frac{1}{z^2}
(\eta_{\mu \nu} dx^\mu dx^\nu - dz^2) \,
, \label{metric}
\end{equation}
where $\eta_{\mu \nu} \! = \! \rm{diag} (+,-,-,-)$ is distinguished
from the full nontrivial 5D metric $g^{\vphantom\dagger}_{MN}$
obtained from Eq.~(\ref{metric}).  One conjectures that weakly-coupled
gravity on this background corresponds to strongly-coupled QCD.
Crudely speaking, the $z$ ``bulk'' coordinate corresponds to an
inverse momentum scale: $z \! \equiv \! \epsilon \! \to \!  0$
corresponds to the UV limit, while $z \! > \! 0$ probes the IR
behavior of the theory.  Every QCD operator ${\cal O}(x)$ is sourced
by a 5D operator $\Psi(x,z)$ that is uniquely determined by its
boundary value $\Psi(x,\epsilon)$, hence the term ``holographic.''
The behavior of $\Psi(x,z)$ for $z \! > \! 0$ then encapsulates the IR
dynamics.  Lastly, the global symmetry of isospin is promoted to a
gauged symmetry in the bulk.

\section{Formalism}

The application of these assertions provides a holographic dictionary
between QCD and the 5D theory.  In particular, to the QCD quark
bilinear operators $\bar q_L \gamma^\mu t^a q_L$, $\bar q_R \gamma^\mu
t^a q_R$, and $\bar q_R^\alpha q_L$, one associates the gauge fields
$A^a_{L\mu}$ and $A^a_{R\mu}$ (with coupling $g_5$), and the
bifundamental field $(2/z)X^{\alpha \beta}$, respectively.  From these
one defines polar- and axial-vector fields $V^M \! , A^M \!
\equiv \! \frac 1 2 (A_L^M \pm A_R^M)$, from which one defines
field strengths $F_V^{MN} \! \equiv \! \partial^M V^N \! - \!
\partial^N V^M \! - i([V^M, V^N] \! + \! [A^M, A^N])$ and $F_A^{MN} \!
\equiv \! \partial^M A^N \! - \! \partial^N A^M \! - i([V^M, A^N] \! +
\! [A^M, V^N])$ and the covariant derivative $D^M X \! \equiv \!
\partial^M X \! - i[V^M, X] \! - i \{ A^M, X \}$.  One further
decomposes $X \! = \! X_0 \exp (2i\pi^a t^a )$, where the modulus
field $X_0 \! = \! \frac 1 2 v(z)$ carries information on the form of
chiral symmetry breaking (as discussed below), and the exponent is the
usual nonlinear representation of pseudoscalar pion fields (in this
notation, we take $\pi \! \equiv \tilde \pi /f_\pi$, where $\tilde
\pi$ is the canonically-normalized pion field and $f_\pi \! = \!
92.1$~MeV).  The lowest-order 5D action then reads
\begin{equation}\label{5DL}
S = \int\! d^{\, 5} \! x \, e^{-\Phi(z)} \, \sqrt{g}\, {\rm Tr}
\left\{ |DX|^2 + 3 |X|^2 - \frac{1}{2g_5^2} (F_V^2 + F_A^2) \right\} \,
,
\end{equation}
where $e^{-\Phi(z)}$ represents a background dilaton coupling.
Working in the axial-like gauges $V_z \! = \! A_z \! = \! 0$,
resolving $A_\mu \! = \!  A_{\mu \perp} \! + \! \partial_\mu \varphi$
into transverse and longitudinal parts, and working in momentum space,
one obtains the Euler-Lagrange equations:
\begin{equation}\label{eqVAdS}
\partial_z\left(\frac{e^{-\Phi(z)}}{z} \, \partial_z V_\mu^a \right)
 + \frac{q^2 e^{-\Phi(z)}}{z} V_\mu^a = 0 \, ,
\end{equation}
\begin{equation}
  \left[ \partial_z\left(\frac{e^{-\Phi(z)}}{z} \, \partial_z A^a_\mu
\right) + \frac{q^2 e^{-\Phi(z)}}{z} A^a_\mu - \frac{g_5^2 \, v(z)^2
e^{-\Phi(z)}}{z^3} A^a_\mu \right]_\perp =0 \, ,
\label{AT}
\end{equation}
\begin{equation}
  \partial_z\left(\frac{e^{-\Phi(z)}}{z} \, \partial_z \varphi^a
\right) +\frac{g_5^2 \, v(z)^2 e^{-\Phi(z)}}{z^3} (\pi^a-\varphi^a) =
0 \, ,
\label{AL}
\end{equation}
\begin{equation}
  -q^2\partial_z\varphi^a+\frac{g_5^2 \, v(z)^2}{z^2} \, \partial_z
\pi^a =0 \, ,
\label{Az}
\end{equation}
\begin{equation}
\partial_z \left( \frac{e^{-\Phi(z)}}{z^3} \partial_z X_0 \right) +
\frac{3e^{-\Phi(z)}}{z^5} X_0 = 0 \, .
\label{Xeqn}
\end{equation}
The meson masses/wave functions are then obtained as the
eigenvalues/eigenfunctions of the equations of motion treated as
Sturm-Liouville systems, leading to Kaluza-Klein towers of meson
states reminiscent of old-fashioned Regge trajectories.  Meanwhile,
the source currents that create or destroy mesons appear as the
free-field solutions to the equations of motion.  Meson form factors
are then overlap integrals (in $z$) of the source solutions with the
external-state eigenmode solutions.  Our case of interest is the pion
electromagnetic form factor $F_\pi(Q^2)$; the pion wave function is
the lowest-mass mode (massless in the limit $m_q \! \to \!  0$) of the
field $\pi (q^2,z)$, which is seen from Eqs.~(\ref{AL})--(\ref{Az}) to
be coupled to the solution for $\varphi(q^2,z)$.  With $V(q,z)$ being
the source current from Eq.~(\ref{eqVAdS}) normalized by $V(0,z) \! =
\!  1$, one obtains
\begin{equation}
F_\pi(q^2) = \int dz \, e^{-\Phi(z)} \,
\frac{V(q,z)}{f_\pi^2} \left\{ \frac{1}{g_5^2 z}
[\partial_z\varphi(z)]^2 + \frac{v(z)^2}{z^3} \left[\pi(z) -
\varphi(z)\right]^2 \right\} \ , \label{ff} 
\end{equation}
which is most useful for spacelike $Q^2$.  In the timelike region, the
large-$N_c$ nature of the holographic approach gives the form factor
as a sum over zero-width vector meson poles (the $\rho$ and its
excitations):
\begin{equation}
 \label{pion_timelike}
 F_\pi(q^2) = -\sum_{n = 1}^{\infty} \frac{f_{n} g_{n\pi\pi}}
 {q^2 - M^2_{n}} \ ,
\end{equation}
where $g_{n\pi\pi}$ is given by
\begin{equation} \label{gnpipi}
g_{n \pi \pi} = \frac{g_5}{f_\pi^2} \int \! dz \, \psi_n (z) \,
e^{-\Phi(z)} \left\{ \frac{1}{g_5^2 z} [\partial_z \varphi (z)]^2 +
\frac{v(z)^2}{z^3} \left[ \pi (z) - \varphi(z) \right]^2 \right\} \,
\ ,
\end{equation}
where $\psi_n$ are the eigenmodes for the vector states.  While
Ref.~\refcite{KL1} considered both $F_\pi (Q^2)$ curves and the
pattern of $g_{n \pi \pi}$ values, in this talk we focus exclusively
on the spacelike pion form factor $F_\pi (Q^2)$ obtained from
Eq.~(\ref{ff}).

\section{Background and Chiral Symmetry-Breaking Fields}

The discussion to this point is just a straightforward application of
the basic holographic scheme.  To continue from this point, however,
one must make two choices: for the field $v(z)$ encapsulating the
chiral symmetry breaking, and for the background field $e^{-\Phi(z)}$.

Addressing first the background field, we note that two popular
choices permeate the literature: the hard-wall model\cite{PS} with
step function $e^{-\Phi(z)} \! = \! H(z_0 \! - \! z)$ and the
soft-wall model\cite{KKSS} with Gaussian $e^{-\Phi(z)} \! = \!
e^{-\kappa^2 z^2}$.  We introduce\cite{KL2} the interpolating
``semi-hard'' option, inspired by the Saxon-Woods model of nuclear
charge density:
\begin{equation} \label{SW}
e^{-\Phi(z)} = \frac{e^{\lambda^2 z_0^2} - 1}{e^{\lambda^2 z_0^2} +
e^{\lambda^2 z^2} - 2} \, ,
\end{equation}
which, like the hard-wall profile, has a drop-off at $z \! = \!  z_0$,
but like the soft-wall profile decreases as $e^{-\lambda^2 z^2}$ for
large $z$.  The hard-wall model was introduced for its simplicity: The
fields simply permeate a fixed distance $z_0$ into the bulk, and the
resulting meson trajectories as a function of excitation quantum
number $n$ scale as $m_n^2 \! \sim \! n^2$.  On the other hand,
semiclassical flux-tube QCD reasoning leads\cite{Shifman:2005zn} to
the conclusion $m_n^2 \! \sim \! n^1$; the soft-wall model was
developed precisely to accommodate this behavior.  Unfortunately,
hard-wall models tend to give more accurate predictions for QCD
observables than soft-wall models; as an example\cite{GR2}, the
experimental value for the ratio $m_\rho^2/f_\rho \! = \! 5.02 \! \pm
\! 0.04$ compares favorably with the hard-wall result 5.55, but rather
poorly with the soft-wall prediction 8.89 (other examples appear in
Table~\ref{Table1}).  Nevertheless, as seen below, the linear
trajectory of the soft-wall model is preserved simply by the
exponential tail of the background, which motivates the hybrid choice
in Eq.~(\ref{SW}).

Turning now to the choice of chiral symmetry breaking represented by
the field $v(z)$, we begin by noting that the two solutions to
Eq.~(\ref{Xeqn}) in the hard wall case are $z^1$ and $z^3$.  Since the
standard gauge/gravity techniques identify the operator source as the
non-normalizable (more singular) solution and the state and associated
vev with the normalizable (less singular) as $z \! \to \! 0$, one
identifies\cite{EKSS} the coefficient of $z^1$ with the quark mass and
$z^3$ with the quark condensate: $v(z) \! = \!  m_q z \! + \! \sigma
z^3$.  In the soft-wall model, the exact solutions turn out to be
Kummer (confluent hypergeometric) functions, which have the
unfortunate feature that only one solution satisfies the appropriate
boundary conditions\cite{KKSS} by vanishing asymptotically as $z \!
\to \! \infty$.  Taken literally, this unique solution would give
an unphysical fixed ratio for $m_q$ to $\sigma$; as argued in
Ref.~\refcite{KKSS}, however, neglected higher-order terms in the
quark potential permit independent coefficients for the $z^1$ and
$z^3$ terms in the low-$z$ expansion of the solution for $v(z)$.
Lacking exact forms for the higher-order terms, one may implement this
fact in two ways: One may note\cite{KL1} that the soft-wall background
$e^{-\kappa^2 z^2}$ suppresses the distinction between the exact
solution for $v(z)$ and $m_q z + \! \sigma z^3$ at large $z$, or one
may choose\cite{KL2} a modified form for $v(z)$ that behaves like $m_q
z + \sigma z^3$ for small $z$ and like the asymptotic form of the
appropriate Kummer function for large $z$.  In either case, one finds
that the numerical solutions for static observables and $F_\pi (Q^2)$
are never better than those of the hard-wall model.

\section{Numerical Solutions}

This brings us to the question of how to solve
Eqs.~(\ref{eqVAdS})--(\ref{Xeqn}) in practice.  In particular,
Eqs.~(\ref{AL}) and (\ref{Az}) are coupled, and the whole set depends
upon three adjustable parameters, $z_0$ or $\kappa$, $m_q$, and
$\sigma$.  Analytic solutions exist only in certain limits\cite{KL1},
particularly, as studied in Ref.~\refcite{GR3}, $m_q \! \to \! 0$.  If
$m_q \! \neq \! 0$, one must resort to a numerical approach, using
standard techniques\cite{NR} such as the ``shooting method'' and
properly convergent numerical integrations to solve the equations.
Such calculations, carried out in hard, soft, and semi-hard
backgrounds, form the core of our work\cite{KL1,KL2}.

\begin{figure}
\psfig{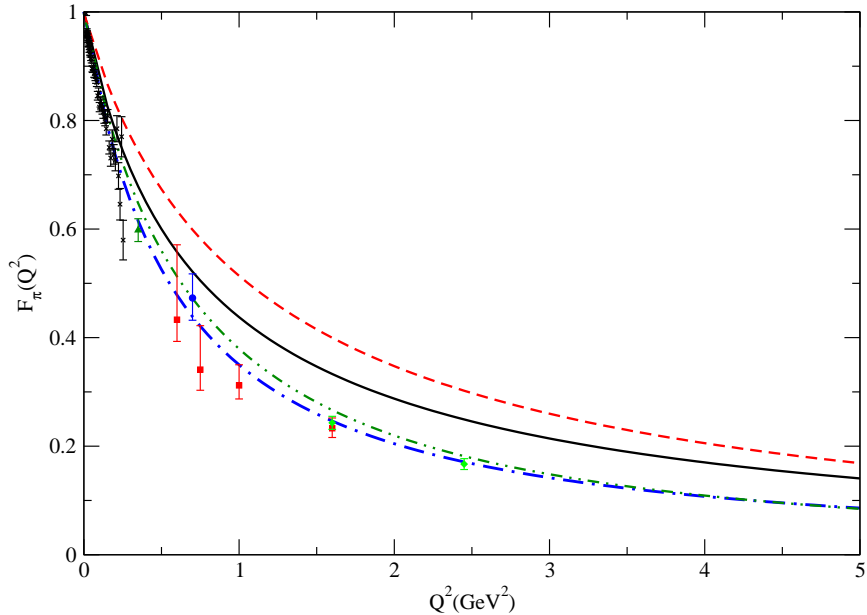}
\caption{The pion form factor $F_\pi(Q^2)$ prediction in hard- and
soft-wall models compared to
data\cite{Ame84,Bra77,Tadevosyan:2007yd,Ack78,Horn:2006tm,bebek}.  The
solid (black online) and dash-dot (blue online) lines are hard-wall
model predictions whose input parameters differ only by use of a
smaller value of $f_\pi$ than experiment in the latter, and
analogously for the dashed (red online) and dash-dot-dot (green
online) lines in the soft-wall model.  The input values appear in the
text.}
\label{ff:fig1}
\end{figure}

\begin{figure}
\psfig{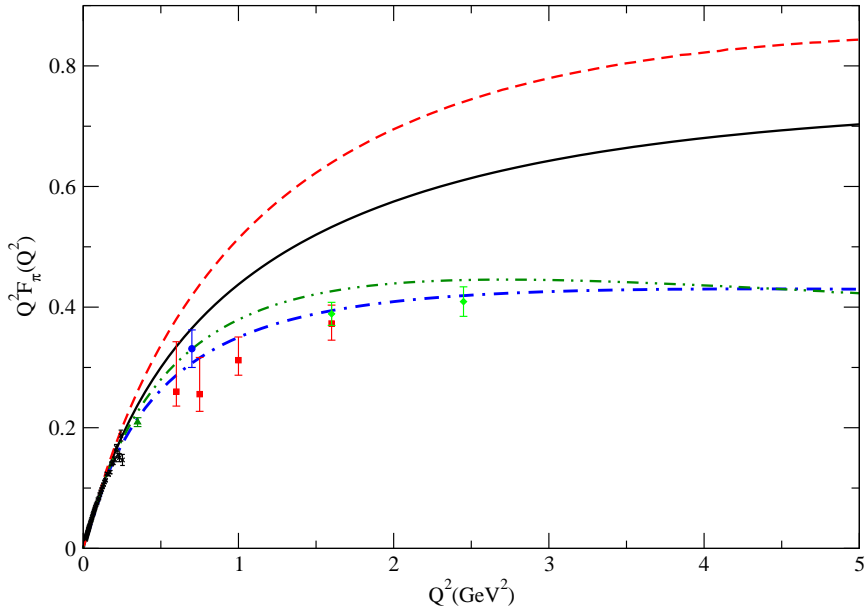}
\caption{The same as Fig.~\ref{ff:fig1}, for the combination $Q^2
F_\pi (Q^2)$.}
\label{ff:fig2}
\end{figure}

Figure~\ref{ff:fig1} compares our $F_\pi(Q^2)$ hard- and soft-wall
model results to data.  The value of $z_0$ (hard wall) or $\kappa$
(soft wall) is completely fixed by the value of $m_\rho$; once this
primary parameter is fixed, the pion decay constant $f_\pi$ is
determined by adjusting $\sigma$.  Finally, the
Gell-Mann--Oakes--Renner formula $m_\pi^2 f_\pi^2 \! = \!  2m_q
\sigma$ uses $m_\pi$ to fix a value of $m_q$.  Empirically, the shape
of $F_\pi (Q^2)$ is driven primarily by $\sigma$.  Using the
experimental values $m_\rho \! = \! 775.3$~MeV, $m_\pi \! = \!
139.6$~MeV, and $f_\pi \! = \! 92.1$~MeV, one obtains the hard-wall
parameters $1/z_0 \! = \! 322$~MeV, $\sigma^{1/3} \! =
\! 326$~MeV, and $m_q \! = \!  2.30$~MeV, which in turn generate the
solid line in Fig.~\ref{ff:fig1}.  The same experimental values for
$m_\rho$ and $m_\pi$ (but taking\footnote{The slightly smaller value
used for $f_\pi$ allows for a much improved fit to $F_\pi(Q^2)$.}
$f_\pi \! = \!  87.0$~MeV) in the soft-wall model give $\kappa \! = \!
389$~MeV, $\sigma^{1/3} \! = \! 368$~MeV, and $m_q \! = \! 1.45$~MeV,
and produce the dashed line.  Both models predict a value of $F_\pi
(Q^2)$ clearly more shallow than data, an effect even more pronounced
when one views the same plot using the dependent variable $Q^2 F_\pi
(Q^2)$ [Fig.~\ref{ff:fig2}].  Interestingly, the discrepancy with
$F_\pi(Q^2)$ data could easily be cured if $f_\pi$ were smaller:
$f_\pi \! = \! 64.2$~MeV ($\sigma^{1/3} \! = \! 254$~MeV) (hard-wall)
gives the dash-dot curve in Fig.~\ref{ff:fig1}, and $f_\pi \! = \!
52.2$~MeV ($\sigma^{1/3} \! = \! 262$~MeV) (soft-wall) gives the
dash-dot-dot curve.  We return presently to the question of the
meaning of these anomalously small $f_\pi$ values.

Figure~\ref{ff:fig2} is also interesting because it seems to suggest
near-asymptotic values for $Q^2 F_\pi (Q^2)$.  For example, if one
adopts the standard abbreviation $s_0 \! \equiv \! 8\pi^2 f_\pi^2 \! =
\! 0.67$~GeV$^2$, then the original hard-wall model appears to approach
at least $Q^2 F_\pi (Q^2) \! \approx \! 1.2 s_0$ as $Q^2 \! \to \!
\infty$.  In fact, the analytic $m_q \! = \! 0$ hard-wall results of
Ref.~\refcite{GR3} for $Q^2 F_\pi (Q^2)$, which appear to conform
closely with our numerical $m_q \! \neq \! 0$ results, predict that
$Q^2 F_\pi (Q^2) \! \to \! s_0$ as $Q^2 \! \to \! \infty$, but also
that $Q^2 F_\pi (Q^2)$ overshoots its asymptote and does not return to
it until values of $Q^2 \! \gg \! 5$~GeV$^2$, at which partonic
effects (absent in this holographic approach) are expected to become
relevant.  Note that the perturbative QCD result\cite{PQCD} for $Q^2
F_\pi (Q^2)$ scales not as a constant, but rather falls off as
$\alpha_s (Q^2) f_\pi^2$.

\begin{figure}
\psfig{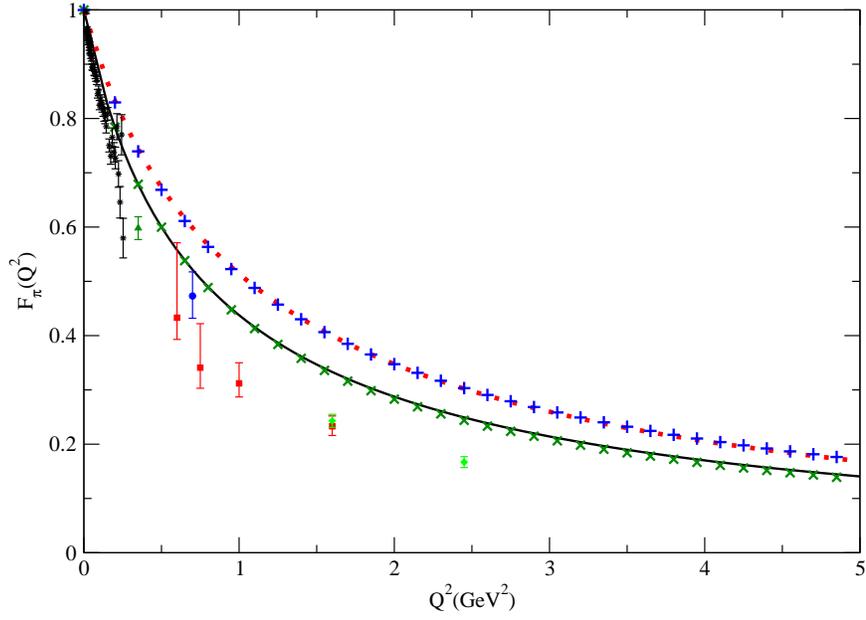}
\caption{Data and hard- and soft-wall model results for $F_\pi (Q^2)$
(same symbols as in Fig.~\ref{ff:fig1}), superimposed with semi-hard
model with $\lambda z_0 \! = \! 2.1$ (crosses, green online) and
$\lambda z_0 \! = \! 1.0$ (pluses, blue online).}
\label{ff:fig3}
\end{figure}

We have argued that the semi-hard background in Eq.~(\ref{SW}), for
suitable values of $\lambda$ (or dimensionless variable $\lambda
z_0$), can be made to simulate either hard-wall or soft-wall
backgrounds.  This effect is illustrated in Fig.~\ref{ff:fig3}, which
again presents the data and original hard- (solid) and soft-wall
(dashed) results from Fig.~\ref{ff:fig1}, superimposed with the result
of the semi-hard model for $\lambda z_0 \! = \! 2.1$ (crosses) and
$\lambda z_0 \! = \!  1.0$ (pluses).  While the agreement between the
semi-hard wall and original hard- and soft-wall models for $F_\pi
(Q^2)$ is impressive, one must check that the meson static observables
(masses, decay constants, {\it etc.}) also agree; this is confirmed by
a glance at Table~\ref{Table1}.  Despite agreeing so well with so many
hard-wall quantities, the semi-hard wall model with $\lambda z_0 \! =
\! 2.1$ nevertheless generates a very different meson trajectory, as
illustrated in Table~\ref{Table2}: One finds that the exponential tail
is sufficient, even for the modest value $\lambda z_0 \! = \! 2.1$, to
turn the hard-wall $m_n^2 \! \sim \! n^2$ trajectory into one that is
$\sim \! n^1$, as a careful examination of the numbers confirms.
Thus, the semi-hard wall model carries all the best features of both
hard- and soft-wall models.

\begin{table}
\tbl{Observables in soft- and hard-wall models compared to those from
the semi-hard model of Eq.~(\ref{SW}) with $\lambda z_0 \!= \!1$ and
$\lambda z_0 \! = \! 2.1$, respectively; values in MeV (except for
$g_{\rho \pi \pi}$, which is dimensionless).}
{
\begin{tabular}{cccccc}
\toprule
Observable & Experiment & Soft-wall & $\lambda z_0 \! = \! 1$ &
Hard-wall & $\lambda z_0 \! = \! 2.1$ \\
\colrule
$m_\pi$           & 139.6$\pm0.0004$ \cite{PDG}      & 139.6 & 139.6 &
139.6 & 139.6 \\
$m_\rho$          & 775.5$\pm 0.4$ \cite{PDG}        & 777.4 & 779.2 &
775.3 & 777.5 \\
$m_{a_1}$         & 1230$\pm 40$ \cite{PDG}          & 1601  & 1596  &
1358  & 1343  \\
$f_\pi$           & 92.4$\pm 0.35$ \cite{PDG}        & 87.0  & 92.0  &
92.1  & 88.0  \\
$f_\rho^{\,1/2}$  & 346.2$\pm1.4$ \cite{Donoghue}    & 261   & 283   &
329   & 325   \\
$f_{a_1}^{\,1/2}$ & 433$\pm13$ \cite{SS,Isgur:1988vm}& 558   & 576   &
463   & 474   \\
$g_{\rho\pi\pi}$  & 6.03$\pm0.07$ \cite{PDG}         & 3.33  & 3.49  &
4.48  & 4.63  \\
\botrule
\end{tabular}
}
\label{Table1}
\end{table}
\begin{table}[t]\vspace{-6pt}
\tbl{Comparison of vector meson masses (in MeV) from the hard-wall
model and the semi-hard wall model using $e^{-\Phi(z)}$ of
Eq.~(\ref{SW}) with $\lambda z_0 \! = \! 2.1$.}  {
\begin{tabular}{cccccc}
\toprule
& $n \! = \! 1$ & $n \! = \! 2$ & $n \! = \! 3$ & $n \! = \! 4$ &
$n \! = \! 5$ \\
\colrule
Hard wall      & 775.6 & 1780.2 & 2790.8 & 3802.8 & 4815.2 \\
Semi-hard wall & 777.5 & 1608.1 & 2226.8 & 2637.5 & 2986.6 \\
\botrule
\end{tabular}
}
\label{Table2}
\end{table}

One mystery remains, namely, why all the models considered here using
only experimental inputs predict curves for $F_\pi (Q^2)$ too shallow
compared to data (predicting, {\it e.g.}, too small a pion charge
radius $\langle r_\pi^2 \rangle$).  As mentioned above, partonic
degrees of freedom have not entered into the holographic calculation
in any essential way; indeed, the only place that the fundamental QCD
gauge theory appears is through matching\cite{EKSS} the vector-current
two-point function calculated both in the 5D theory and in
(perturbative) QCD, from which one determines the 5D gauge coupling
$g_5 \! = \! 2\pi$.  It is natural to suppose that the holographic
model is at best incomplete, due to the absence of partonic degrees of
freedom; however, the accurate $F_\pi (Q^2)$ data\footnote{Some $F_\pi
(Q^2)$ data points extend out to about 10~GeV$^2$, but the
uncertainties are sufficiently large as to accommodate almost any
model.} extends only out to $\sim \!  3$~GeV$^2$, far below the regime
where one would expect the partonic expression for $F_\pi (Q^2)$ to
dominate.  Alternately, one might argue that the treatment of chiral
symmetry used here is inadequate; while this is certainly possible, it
appears to be the most realistic treatment available.  These were the
possible explanations proffered in Ref.~\refcite{KL2}.

However, we present here one further possible explanation: The
holographic method implicitly assumes both large $N_c$ and large
't~Hooft coupling $g_s^2 N_c$.  Can the shallow slope of $F_\pi (Q^2)$
be a $1/N_c$ correction?\footnote{That the discrepancy can be a
$1/g_s^2 N_c$ correction was suggested to us by O.~Andreev.}  To test
this idea, we ask which quantities in the analysis are most sensitive
to $1/N_c$ corrections.  While meson masses $m_\pi$ and $m_\rho$ are
$O(N_c^0)$, their decay constants are $O(N_c^{1/2})$ and thus are
fractionally more sensitive to variations in $N_c$.  Recall that the
original hard-wall model gave a perfect account of the $F_\pi(Q^2)$
data if $f_\pi$ had been 64.2~MeV.  This number is, interestingly,
very close to a factor $1 \! - \! 1/N_c$ smaller than the experimental
value 92.1~MeV, so that their difference is easily attributable as a
$1/N_c$ correction.

While such an explanation may seem a bit glib, a similar effect has
been seen long ago in the literature: The pioneering soliton model
work of Adkins, Nappi, and Witten\cite{Adkins:1983ya}.  Using our
normalization for $f_\pi$, their model values inserted into the
Goldberger--Treiman relation $f_\pi \! = \!  M_N g_A / g_{\pi \pi N}$
predict\footnote{Reference~\refcite{Adkins:1983ya} actually used the
Goldberger--Treiman relation to predict $g_{\pi \pi N}$.} $f_\pi \! =
\! 61$~MeV, a result traditionally attributed to being due to a
$1/N_c$ correction.

These considerations suggest a supposedly ``perfect'' holographic
model for low-energy hadronic phenomena that should include two
features: A semi-hard wall background with an exponential tail
extending to large $z$, and $1/N_c$ corrections of a natural size,
particularly for $f_\pi$.  The semi-hard wall model has been seen to
give a fit to low-energy observables just as good as that of the
hard-wall model, but nevertheless generates the desired linear
trajectories for excited mesons.  Meanwhile, allowing $f_\pi$ to be
smaller by a relative $1/N_c$ correction is sufficient to correct the
shallowness of the $F_\pi(Q^2)$ curves compared to experiment.
Indeed, it is remarkable that these modifications are sufficient to
cure the discrepancies with data but still make no mention of the
partonic degrees of freedom in QCD.  A truly perfect holographic model
would of necessity incorporate dynamical quarks as well.

\section{Conclusions}

Since their inception, holographic methods have provided a compelling
theoretical framework in which to study hadronic quantities, including
$F_\pi (Q^2)$.  In this talk we have seen that the choice of
background field behavior has a strong effect on low-energy
observables, but it is possible to retain many of the best features of
each model while adjusting this background; the semi-hard wall
background proves ideal for accomplishing this goal.  In passing, we
note that the treatment of chiral symmetry breaking advocated by
Ref.~\refcite{EKSS} appears completely suitable for this purpose.
However, we begin to see in the precise value of the slope of
$F_\pi(Q^2)$ possible evidence for the necessity of including $1/N_c$
corrections in order to achieve completely satisfactory agreement with
data.  The great remaining challenge appears to be how to knit
together the promising first results of this all-hadronic approach
with the fundamental QCD partonic degrees of freedom.

\section*{Acknowledgments}
RFL thanks the organizers for their kind invitation and an interesting
scientific program.  This presentation benefited from discussions with
H.~Grigoryan and A.~Radyushkin.  This work was supported by the NSF
under Grant No.\ PHY-0456520.

\end{document}